# Search for scaling dimensions for random surfaces with c=1

by

*J. Ambjørn* [1], *P. Białas* [2], *Z. Burda* [3,4,5], *J. Jurkiewicz* [1,5] *and B. Petersson* [3]

## Abstract

We study numerically the fractal structure of the intrinsic geometry of random surfaces coupled to matter fields with $c = 1$. Using baby universe surgery it was possible to simulate randomly triangulated surfaces made of 260.000 triangles. Our results are consistent with the theoretical prediction $d_H = 2 + \sqrt{2}$ for the intrinsic Hausdorff dimension.

[1]The Niels Bohr Institute, Blegdamsvej 17, DK-2100 Copenhagen Ø, Denmark
[2]Institute of Computer Science, Jagellonian University, ul. Nawojki 11, 30-072 Kraków, Poland
[3]Fakultät für Physik, Universität Bielefeld, Postfach 10 01 31, Bielefeld 33501, Germany
[4]A fellow of the Alexander von Humboldt Foundation.
[5]Permanent address: Institute of Physics, Jagellonian University, ul. Reymonta 4, PL-30 059, Kraków 16, Poland

# Introduction

During the last years great progress has been made in the understanding of two dimensional quantum gravity (see [1] for a review, and references therein). Both the continuum and discrete approaches resulted in a consistent description of the universal critical properties of 2d gravity interacting with conformal matter fields with $c \leq 1$. As a consequence, an effective picture of a typical quantum surface of 2d gravity was obtained in terms of self similar branching structure of baby universes which results directly from the scaling of surface entropy, and vice versa [2]. Despite the progress made, there are still some important issues requiring elaboration. One of them is the description of the intrinsic geometry by its internal scaling dimensions.

In the dynamical triangulation approach, the surface is built from equilateral triangles glued along the edges. Each surface has a dual $\phi^3$ graph associated with it, with vertices corresponding to the centers of the triangles of the surface. One can define the intrinsic geometry using the concept of a geodesic distance between any two vertices on a graph. In the dual formulation we define a distance $r$ between two triangles as the length of the shortest path connecting the vertices of the dual graph following the links of the graph. Similarly one can define the distance $r^*$ between the vertices of the surface as the length of the shortest path along the edges of triangles. On a surface built from the equilateral triangles the length of geodesic is just the number of links of the path. These two definitions differ for any particular surface, but we expect the scaling properties to be the same in the ensemble of surfaces.

Let $v(R)$ be a number of points whose geodesic distance from a certain reference point, $p$ on a dual graph is less or equal $R$. These points form a ball (or disc) with a radius $R$. One defines internal Hausdorff dimension $d_H$ by:

$$\langle v(R) \rangle \longrightarrow R^{d_H} \qquad \text{for } R \to \infty \tag{1}$$

The averaging $\langle ... \rangle$ goes over the points on a graph and over different graphs contributing to the ensemble. Because of the surface branching, a typical disc is not simply–connected and hence its boundary is disconnected. Denote the number of connected parts of the boundary by $n(R)$ and define the so called branching scaling dimension $d_B$ by :

$$\langle n(R) \rangle \longrightarrow R^{d_B} \qquad \text{for } R \to \infty \tag{2}$$

The values of scaling dimensions are known, $d_H = 2$ and $d_B = 1$, for random surfaces with large positive $c$ which correspond to branched polymers [3], and for surfaces with large negative $c$, which are flat, $d_H = 2$ and $d_B = 0$. The dimensions are, however, not known in general.

The definition (1) of the scaling dimension is sometimes referred to as a "mathematical" definition to distinguish from the "physical" definition based on the "physical" distance defined from the asymptotic fall–off of a massive propagator on a graph [4]. In [5] it was shown that the massive propagator distance and the mathematical distance, and hence the Hausdorff dimensions, are equivalent.

Numerical measurements of the scaling dimensions require simulation of large lattices. The range of distances – the discs' radii $R$, in which the formula (1) is applicable, is on one hand limited from below to distances much larger than the cut-off, but on the other hand the discs must be much smaller that the system itself. This can be written as $1 \ll R \ll \langle \bar{R} \rangle$, where $\langle \bar{R} \rangle$ is the average geodesic separation between points in the system, which provides a kind of typical scale for linear extension of the system. The range limitation for $R$ make a direct use of the definition of scaling dimension difficult and usually one cannot avoid strong finite size effects. Therefore we propose for numerical purposes to consider an alternative definition of the Hausdorff dimension. For a given ensemble with the area $N_t$ (number of of triangles) we find an average geodesic separation between points and then look how it scales with the lattice size. We define, the scaling dimension $D_H$ by

$$\langle \bar{R} \rangle \longrightarrow \mathrm{N}_t^{1/D_H} \tag{3}$$

and use a capital letter to distinguish it from $d_H$. In the infinite size limit we expect $D_H = d_H$. The advantage of using the definition (3) is that finite size effects are expected to be smaller.

The largest lattices used until now in measurements of the scaling dimensions were of the size $1.3 \cdot 10^5$ triangles in pure gravity [6] and $5.0 \cdot 10^6$ triangles in the case of gravity interacting with matter fields with $c = -2$ [7]. It was possible to reach these large lattice sizes due to the recursive sampling technique specific for the two cases. Recently we have proposed baby universe surgery as a suitable algorithm for simulation of the large systems in general $2d$ quantum gravity systems [8]. In the present paper we use this algorithm to simulate random surfaces coupled to a single gaussian field, i.e. a matter system with $c = 1$. The field $X_i$ is concentrated in the center of the triangle $i$ and the action of the gaussian field is $S = \sum_{ij}(X_i - X_j)^2$, where the sum runs over the links of the dual lattice (pairs $ij$ of neighboring triangles). We simulate lattices up to the size of $2.6 \cdot 10^5$ triangles.

# The Algorithm

The algorithm has been presented in detail in the paper [8]. The main idea is to use underlying branching structure of a typical surface in the update scheme. The basic concept is that of a minimal neck, defined as a loop of length three which does not form a triangle of the surface. Such a loop cuts the surface in two parts, the smaller of which is called a minimal loop baby universe, abbreviated *minbu*. The elementary step of the algorithm is to find a minbu on surface, then cut it and paste in place of a randomly chosen triangle. To assure detailed balance special care should be taken when choosing the minbus. They can be selected with a uniform distribution as was assumed in [8]. Here we use slightly different version of the algorithm. To pick up a minbu, we chose randomly a link on the surface and find all minimal necks which contain this link. If the number of the necks is zero, we chose another link. In case there are minimal necks, the number of them can be larger than 1. Denote by $n_{old}$ this number for a given link $l_{old}$. We select one of the minimal necks with probability $1/n_{old}$. The minbu is then pasted into a triangle lying outside the minbu. There are six ways of pasting a minbu on a triangle corresponding to six link permutations. In the algorithm we choose randomly one of them. Thus the position of the link $l_{old}$ is known exactly after the cut/paste operation. Let it be $l_{new}$. Denote by $n_{new}$ the number of the minimal necks which will contain the link $l_{new}$ after the paste operation. In the inverse cut/paste operation the minimal neck corresponding to the newly pasted minbu would be chosen with probability $1/n_{new}$ from the set of minimal loops containing the link $l_{new}$. As a consequence in this link–oriented algorithm we need an additional factor $n_{new}/n_{old}$ in the transition probability to assure that the detailed balance condition is satisfied. This factor was absent in the case when the minimal loops were picked up uniformly.

| $N_t$ | $r^2$ |
|---|---|
| 60 | 0.45482(55) |
| 124 | 0.60107(73) |
| 508 | 0.947(18) |
| 1024 | 1.158(27) |
| 1596 | 1.294(37) |
| 2048 | 1.376(4) |
| 4096 | 1.612(5) |
| 8192 | 1.87(13) |
| 16384 | 2.16(21) |
| 65536 | 2.73(51) |
| 262144 | 3.8(39) |

Table 1:

To test the new version, apart from the basic checks relying on comparison with the standard algorithm for small lattices, we looked at the average action and the gyration radius. The gaussian action per link is : $\langle\sum_{ij}(X_i - X_j)^2\rangle/N_l = (1 - 1/N_t)/3$, where the sum runs over the pairs of triangles and $N_l$, $N_t$ are the numbers of links and triangles, respectively. The data points fitted to this formula give $\chi^2$/d.o.f. = 1.4. In table 1 we have gathered the measured values for the gyration radius and fitted to the theoretical formula $\overline{\langle(X - \bar{X})^2\rangle} = a + b \log N_t + c(\log N_t)^2$. We obtain the following values for the fit parameters : $a = 0.047(37)$, $b = 0.013(16)$, $c = 0.0211(18)$ and $\chi^2$/d.o.f. = 1.09 which can be compared with those found in [9]. The value of $c$ cited there agrees within the errors with our value. The authors [9] did not give the values of $a$ and $b$. We checked, however, that our fit goes through the data points displayed in a figure in that work. Notice, that in the mentioned paper [9] both the action and the definition of gyration radius differ by a factor 2 from the ones used here. This altogether yields the factor 4 between the resulting values of the gyration radius.

During the cut/paste operation the change in the gaussian field is chosen to maximize the acceptance rate. This is achieved by a heat–bath–like algorithm [8]. The cut/paste operations are mixed with flips and shifts of the standard algorithm.

## Measurements

The measurements of intrinsic geometry were performed mainly on graphs dual to the triangulations. To find the number of points $l_p(R)$ lying at a distance $R$ from a given point $p$ on a graph, we follow the standard technique. We find a layer of points at a distance one, by visiting all its neighbors. In our case there are always 3 points since it is a $\phi^3$ graph. Next we visit all the neighbours of the points in the first layer, which have not yet been visited to determine the set of points at a distance 2. Repeating recursively the procedure we fill the histogram $l_p(R)$ until the whole graph gets covered. Integrating over distances we obtain the number of points within a distance $R$, i.e. the disc area $v_p(R) = \sum_{r=1}^{R} l_p(r)$. The easiest way to determine the number of disconnected parts of the disc boundary is to find in addition the numbers of links and elementary faces on the disc and make use of Euler's formula $n_p(R) = 2 + \#links - \#point - \#faces$. Averaging over points $p$ and different surfaces of the ensemble gives us quantities $\langle v(R)\rangle$ and $\langle n(R)\rangle$ on the lhs of (1) and (2).

The distributions $l_p(R)$ can be used to define the average geodesic separation between points on the surface :

$$\bar{R} = \left\langle \frac{\sum_R R \cdot l_p(R)}{\sum_R l_p(R)} \right\rangle_p . \tag{4}$$

In a similar way we can define higher moments $\overline{R^p}$ of the distribution $l_p(R)$. The averaging $\langle..\rangle_p$ is over all points on a graph. In practice it is approximated by sampling over a certain random subset of them since otherwise the measurement would be to time consuming.

We performed simulations for lattices with the sizes being a power of 2, ranging from $2^{10}(1024)$ to $2^{18}$ (262144) triangles, skipping $2^{15}$ and $2^{17}$. For smaller lattices (up to $2^{14}$) we collected the data during more than $10^4$ integrated correlation times $\tau$ of the slowest mode in the update scheme. For small lattices the slowest mode corresponds to the gyration radius, $\overline{(X-\bar{X})^2}$, while for larger lattices it is the average geodesic distance $\bar{R}$ [8]. The measurements were then taken every $10\tau$. The distribution of $n_p(R)$ and $l_p(R)$ were averaged over samples of $10^4$ points $p$. We discarded the data from the first $100\tau$ of $\overline{(X-\bar{X})^2}$. For the largest lattice we run the simulation for around $60\tau$ taking measurements roughly every autocorrelation time. The thermalization took $20\tau$. The computations were done on the HP 715/720 workstations and on the SG Challenge L workstation. The presented data require the equivalent of 3 months of CPU time on HP 720.

## Results

Let us first discuss the branching dimension $d_B$ (2). In fig.1a we plot in the log log scale the number of boundaries $\langle n(R)\rangle$ against the radius $R$ for different lattice sizes. The decrease of $\langle n(R)\rangle$ for $R \sim \langle R\rangle$ comes from the finite size of the lattices. The smaller $R$ is, the weaker the finite size effects affect the distribution $\langle n(R)\rangle$. For triangulations large enough, $R$ can be taken much larger than 1 and much smaller than $\langle R\rangle$ so that one can simultaneously investigate features of the continuous geometry and is not affected by the finite size of the system. This is the region where we look for the scaling dimension. For different lattice sizes we define numerically the quantity $\mathrm{d}\ln\langle n(R)\rangle/\mathrm{d}\ln R$ and check whether it goes to a constant in a certain range $1 \ll R \ll \langle\bar{R}\rangle$ in the limit of large sizes. The results of differentiating is shown in fig.1b. Comparing the curves for two largest lattices, one can see the first indication of the occurrence of an interval in $R$ in which the dependence of $d\ln\langle n(R)\rangle/d\ln R$ on the lattice size develops a plateau. This is more transparently seen for the same quantity measured on the original surfaces. The latter quantity is presented in fig.2b which is the logarithmic derivative of the number of boundaries $\langle n^*(R^*)\rangle$ measured for discs on the original graphs (triangulations), shown in fig.2a. The derivative $\mathrm{d}\ln n^*(R^*)/\mathrm{d}\ln R^*$ seems to saturate above 2.5 at a value which is larger than the maximal value of $\mathrm{d}\ln n(R)/\mathrm{d}\ln R$ found on the dual graphs. This is a result of different finite size effects on the triangulations and the dual graphs. One believes however that in the infinite volume limit they will approach the same universal value reflecting universal, regularization independent continuum geometry. To check this explicitly one should go to even larger lattices. In fig.3a,b we plot the disc volume $v(R)$ and its logarithmic derivative $\mathrm{d}\ln v(R)/\mathrm{d}\ln R$ versus $R$. Similarly to the case $c = -2$ [7], the height of the maxima of the derivative $\mathrm{d}\ln v(R)/\mathrm{d}\ln R$ saturates much more slowly with $R$ than was the case for $\mathrm{d}\ln n(R)/\mathrm{d}\ln R$. In fig.4a we plot in diamonds the heights of the maxima of $\mathrm{d}\ln v(R)/\mathrm{d}\ln R$ for different lattice sizes. In the same picture we plot in solid line the values of the effective Hausdorff dimension $D_H$ obtained from the formula (3). More precisely, the curve is obtained by taking a derivative of the curve going through the data points for the average separation $\langle\bar{R}\rangle$ versus lattice size $N_t$ presented in fig.4b. In order to take the derivative we first fitted an auxiliary function (a third order polynomial in the log log variable) to the data points in fig.4b, and then we took its derivative $\mathrm{d}\ln\langle\bar{R}\rangle/\mathrm{d}\ln N_t$. It's inverse is plotted in fig.4a. The polynomial suffices for this purpose in the sense that it gives a smooth function (see dashed line in fig.4b) with high fit goodness : $\chi^2$/d.o.f. around 1. Needless to say one cannot take this fit as an extrapolation formula for the effective Hausdorff dimension $D_H$.

The curves representing the effective Hausdorff dimension $d_H$ and $D_H$ show gradual flattening. Though it is difficult to make extrapolations of the curves it is tempting to say that both curves approach the same asymptotic value compatible with the prediction $2 + \sqrt{2}$ [10].

# Discussion

The results presented in this paper indicate the existence of the internal scaling dimensions for random surfaces with $c = 1$. The value of the branching dimension can be roughly estimated to be around $d_B = 2.5$ which seem to be very close to that for $c = -2$ and is much higher than for branched polymers (corresponding to $c = \infty$). It would be very interesting to investigate how it changes with $c$ just above the $c = 1$ barrier.

From our data we can put a lower bound on the Hausdorff dimension: $d_H > 3$. The effective value seems to tend to the theoretical prediction $2 + \sqrt{2}$ [10], but because we do not know a finite size extrapolation formula we cannot rule out the value $2+2\sqrt{2}$ advocated in the paper [11]. At present it cannot be ruled out either that the fermionic [10] and diffusion [11] definitions give a different Hausdorff dimension which, as suggested in [4] can be definition dependent. We hope to study this question in the future.

## Acknowledgments

We acknowledge the High Energy Physics Department INP Cracow for the computer time on SG Challenge L, on which a part of the computation was done. P.B. thanks the University of Bielefeld for the kind hospitality during his stay there. Z.B. wishes to thank Alexander von Humboldt Foundation for the fellowship. This work was partially supported by the KBN grant 2P30204705.

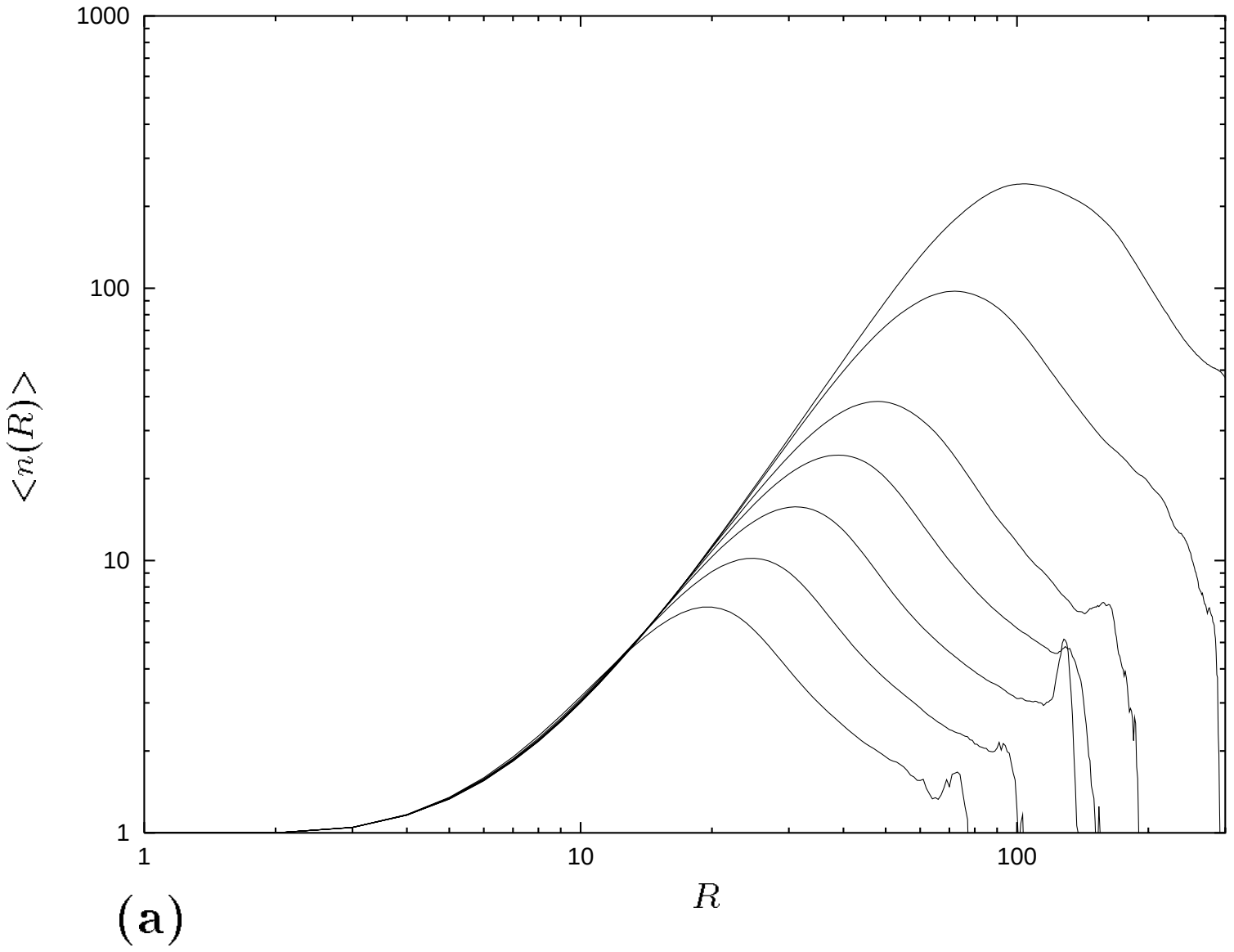

**(a)**

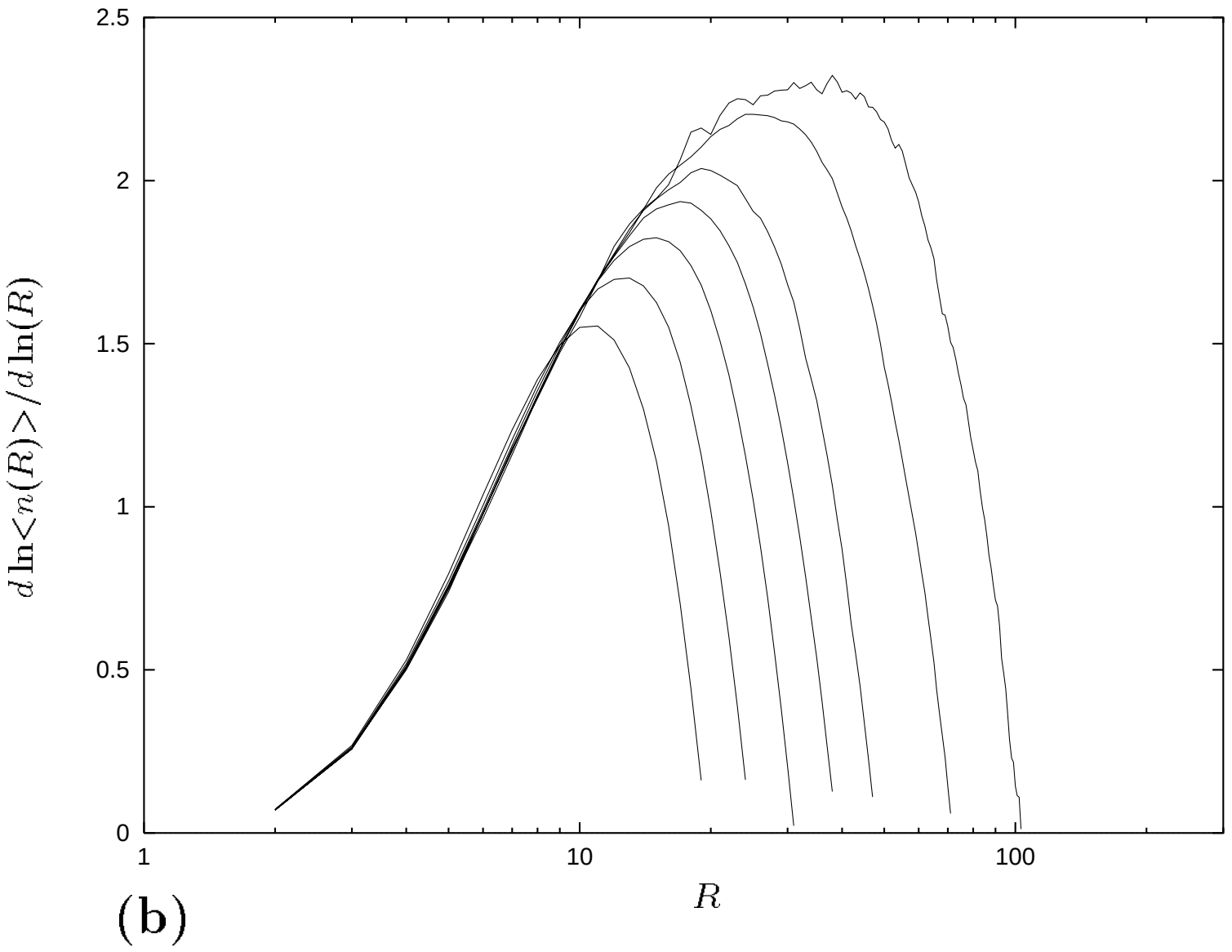

**(b)**

# Figure 1

**(a)** The distributions of the numbers of boundaries, $n(R)$, for the lattice sizes $2^{10}, \ldots, x2^{14}$, $2^{16}$ and $2^{18}$ plotted in log log scale.
**(b)** The logarithmic derivative $\mathrm{d} \ln n(R)/\mathrm{d} \ln R$ of the distributions from the figure (a).

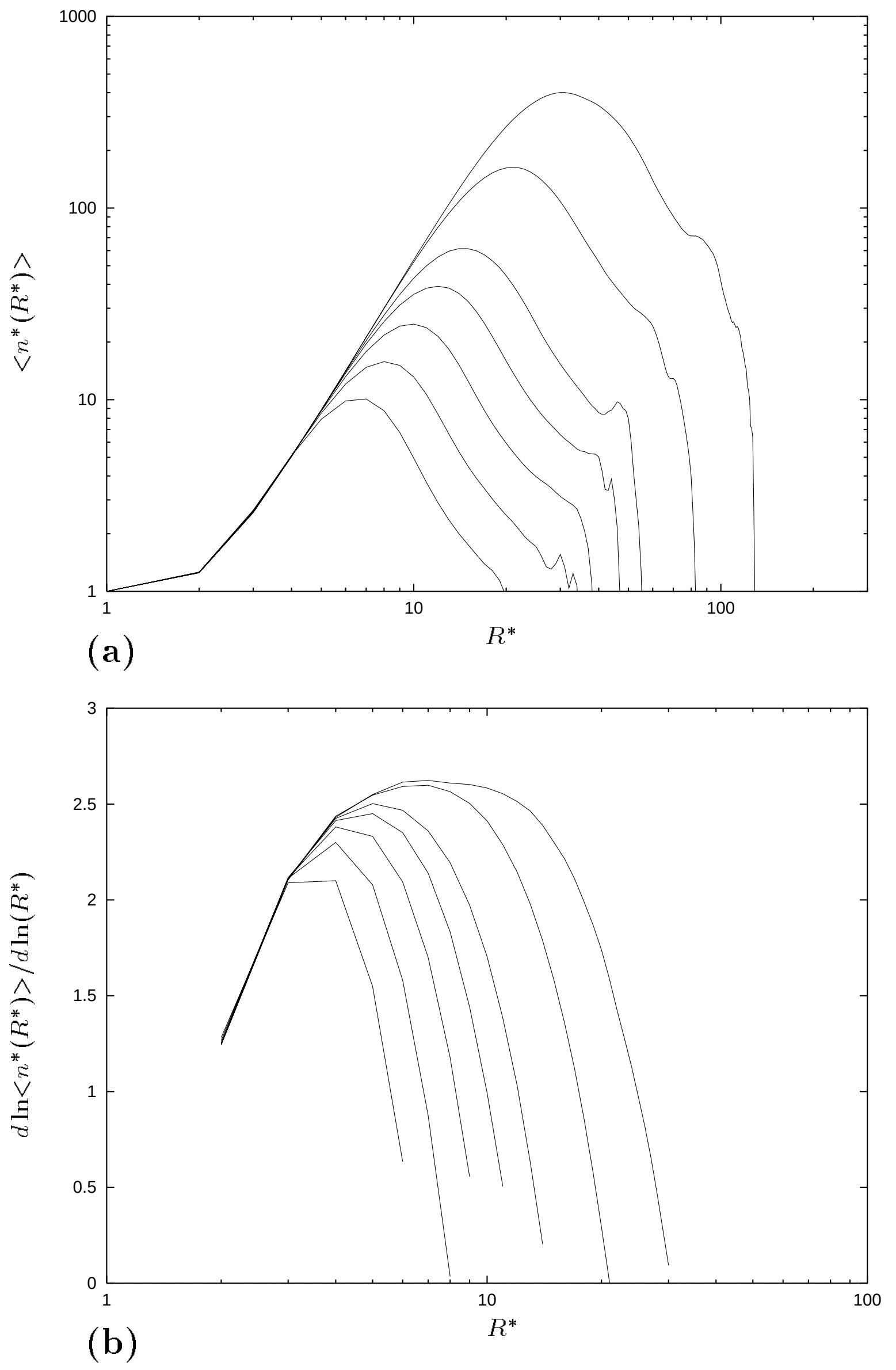

Figure 2

**(a)** The distributions of the numbers of boundaries, $n^*(R^*)$, for the lattice sizes $2^{10}, \ldots, 2^{14}$, $2^{16}$ and $2^{18}$ plotted in log log scale.
**(b)** The logarithmic derivative $\mathrm{d}\ln n^*(R^*)/\mathrm{d}\ln R^*$ of the distributions from the figure (a).

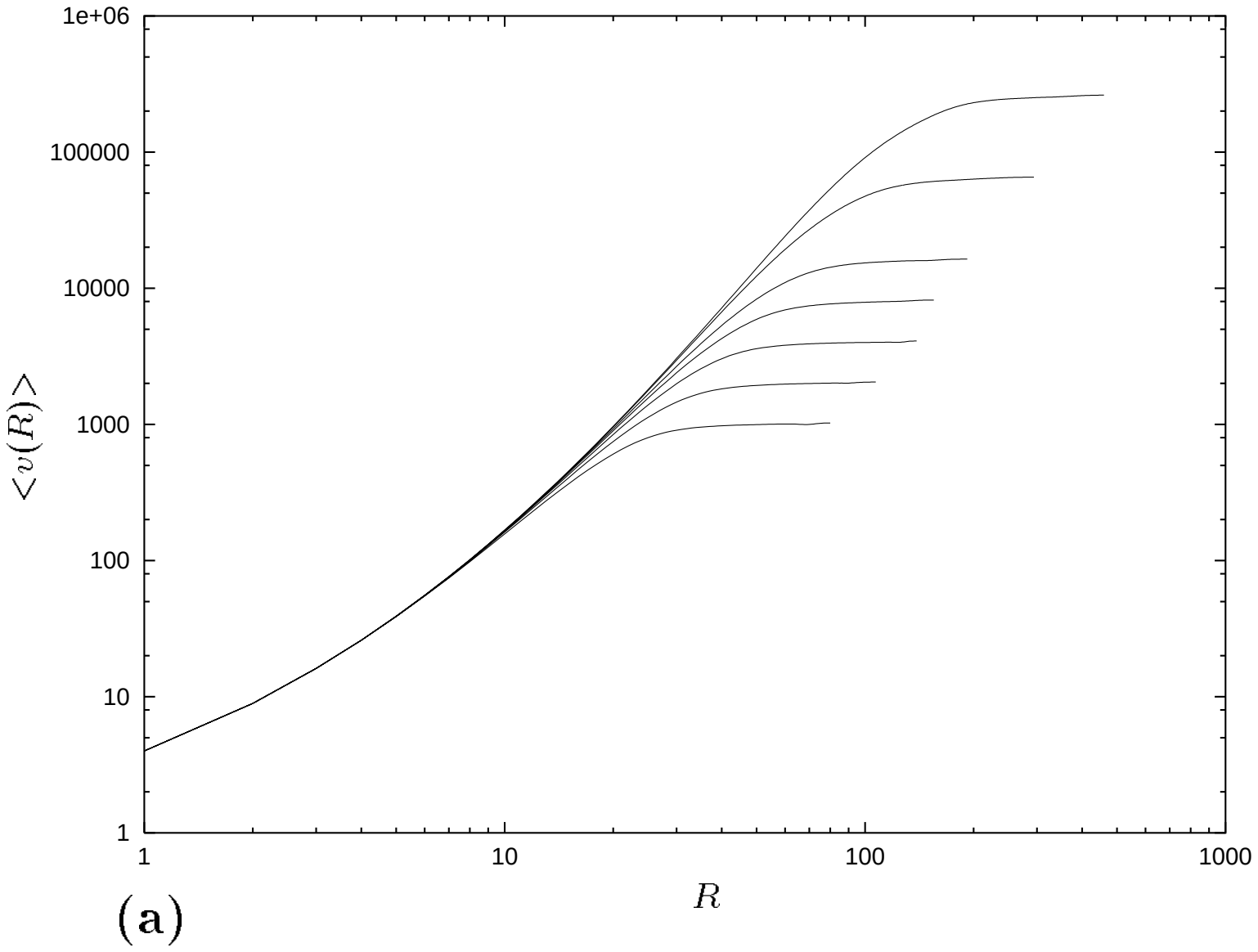

**(a)**

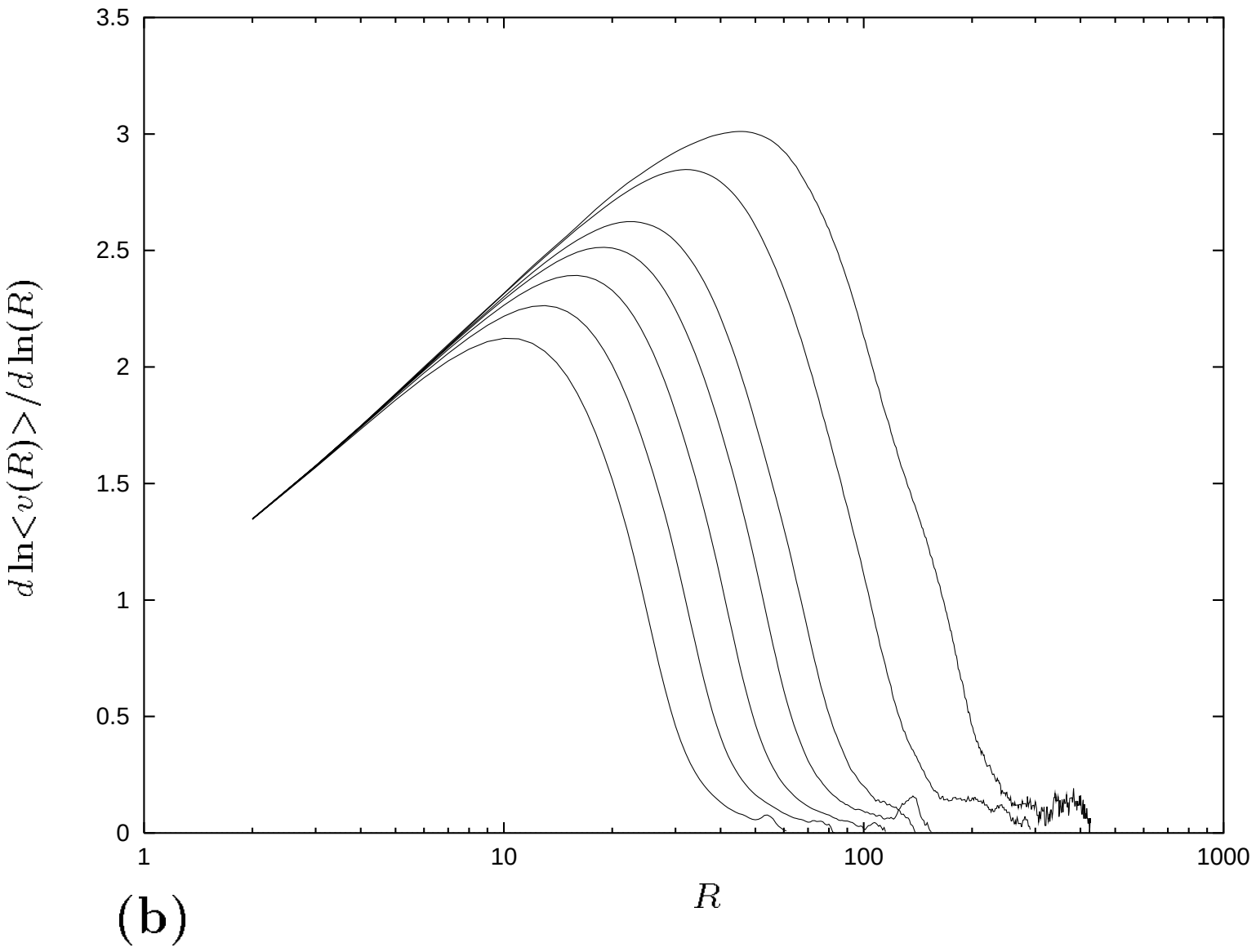

**(b)**

# Figure 3

**(a)** The disc volume, $v(R)$, as a function of the radius for $n(R)$ for the lattice sizes $2^{10},\ldots,2^{14}$, $2^{16}$ and $2^{18}$ plotted in log log scale.
**(b)** The logarithmic derivative $\mathrm{d}\ln v(R)/\mathrm{d}\ln R$ of the distributions from the figure (a).

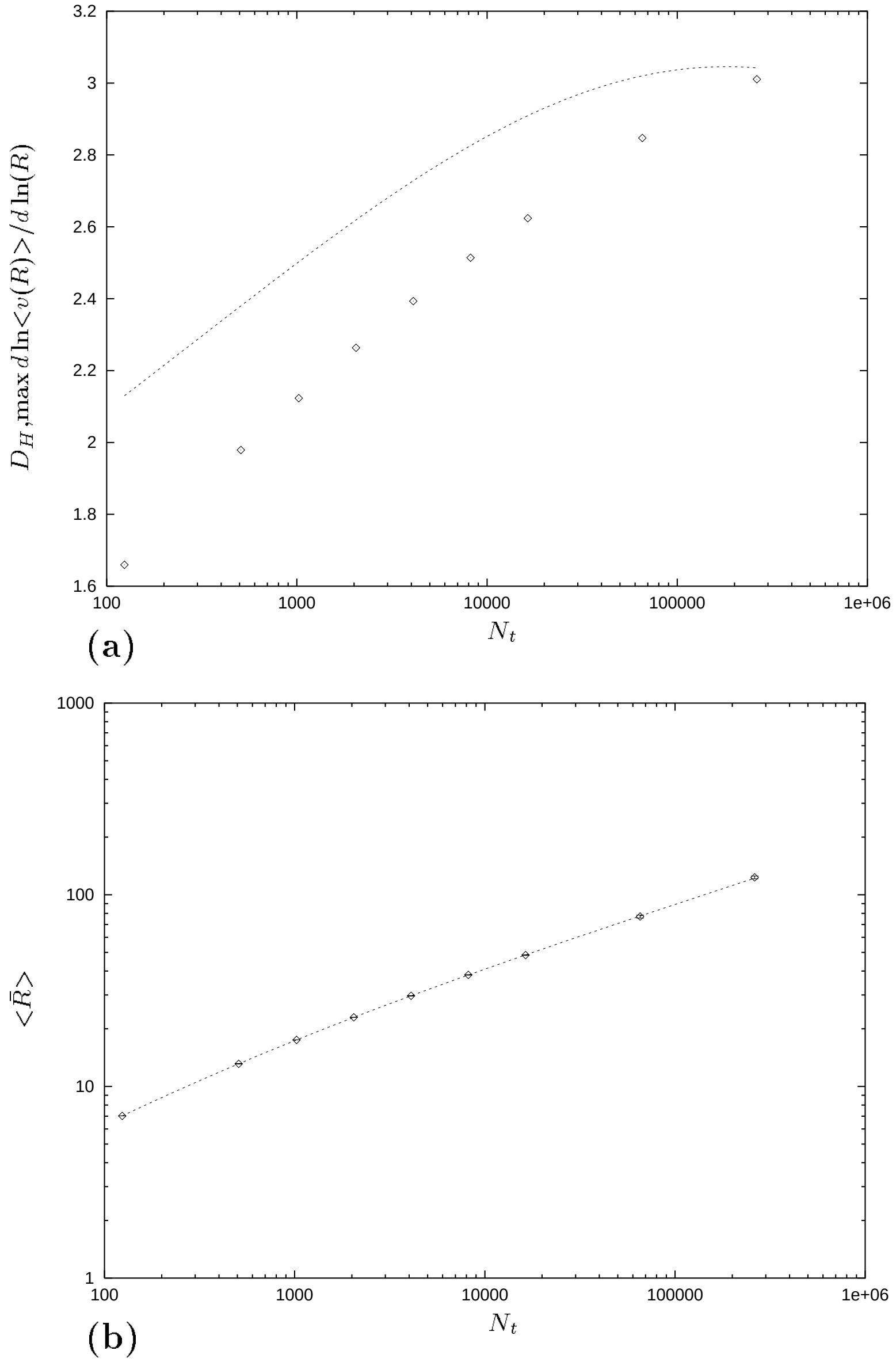

Figure 4

**(a)** The position of the maxima of the logarithmic derivative $\mathrm{d}\ln v(R)/\mathrm{d}\ln R$ (fig.3b) for different lattice sizes ($\diamond$). The effective Hausdorff dimension $D_H$ obtained from the logarithmic derivative $\mathrm{d}\ln\langle\bar{R}\rangle/\mathrm{d}\ln N_t$ of the fit presented in fig. **(b)**.